\documentclass[runningheads]{llncs}
\usepackage{makeidx}  
\usepackage{booktabs}
\usepackage{algpseudocode} 
\usepackage{framed}
\usepackage{color,graphicx}
\usepackage{multirow}
\usepackage{array}
\usepackage{float} 
\usepackage{subfig}
\def\iname#1#2{{$#1_{#2}$}{}}

\begin{document}
\title{Efficient Characterization of Hidden Processor Memory Hierarchies} 
%
%
\author{Keith Cooper\orcidID{0000-0003-4288-4847} \and Xiaoran Xu\orcidID{0000-0003-4450-8757} 
}
\authorrunning{K. Cooper and X. Xu}
%
\institute{Rice University, Houston TX 77005, USA \\
\email{\{keith, xiaoran.xu\}@rice.edu}}
\maketitle              
\begin{abstract}
 A processor's memory hierarchy has a major impact on the performance of running code. 
However,
computing platforms, where the actual hardware characteristics are hidden from both the end user and the tools that mediate execution, such as a compiler, a JIT and a runtime system, are used more and more, for example, performing large scale computation in cloud and cluster. Even worse, in such environments, a single computation may use a collection of processors with dissimilar characteristics. Ignorance of the performance-critical parameters of the underlying system makes it difficult to improve performance by optimizing the code or adjusting runtime-system behaviors; it also makes application performance harder to understand.

To address this problem, we have developed a suite of portable tools that can efficiently derive many of the parameters of processor memory hierarchies, such as levels, \textit{effective capacity} and latency of caches and TLBs, in a matter of seconds. The tools use a series of carefully considered experiments to produce and analyze cache response curves automatically. The tools are inexpensive enough to be used in a variety of contexts that may include install time, compile time or runtime adaption, or performance understanding tools.
\keywords{Efficient Characterization \and Hidden Memory Hierarchies \and Code Performance \and Portable Tool}
\end{abstract}
\section{Introduction and Motivation}\label{introduction}
Application performance on modern multi-core processors depends heavily on the performance of the system's underlying memory hierarchy. The academic community has a history of developing techniques to measure various parameters on memory hierarchies~\cite{saavedra1995,mcvoy1996,dongarra2004,yotov2005,yotov2005a,duchateau2008,gonzalez2010,sandoval2011,taylor2010,sussman2011,abel2012measurement,gonzalez2013s,casas2014active,casas2016evaluation}. However, as deep, complex, and shared structures have been introduced into memory hierarchies, it has become more difficult to find accurate and detailed information of performance-related characteristics of those hierarchies. 
What's more, to an increasing degree, large-scale computations are performed on platforms where the actual characteristics of the underlying hardware are hidden.
Knowledge of the performance-critical parameters of the underlying hardware can be useful for improving compiled code, adjusting runtime system behaviors, or understanding performance issues.
Specifically, to achieve the best performance on such platforms, the code must be tailored to the detailed memory structure of the target processor. That structure varies widely across different architectures, even for models of the same instruction set architecture (\textsc{isa}). Thus, performance can be limited by the compiler's ability to understand model-specific differences in the memory hierarchy, to tailor the program's behavior, and to adjust runtime behaviors accordingly. At present, few compilers attempt aggressive model-specific optimization. One impediment to development of such compilers is the difficulty of understanding the relevant performance parameters of the target processor.
While manufacturers provide some methods to discover such
information, such as Intel's \textsc{cpuid}, those mechanisms are not standardized across manufacturers or across architectures. A standard assumption is that such data can be found in manuals; in truth, details such as the latency of an L1 TLB miss on an Intel Core i7 processor are rarely listed. What is worse, even the listed information may differ from what a compiler really needs for code optimization, such as full hardware capacity vs. \textit{effective capacity}.

\textit{\textbf{Effective capacity}} is defined as the amount of memory at each level that an application can use before the access latency begins to rise. The effective value for a parameter can be considered an upper bound on the usable fraction of the physical resource. Several authors have advocated the use of effective capacities rather than physical capacities~\cite{moyer1991,qasem2006,luk2001}. In the best case, \textit{effective capacity} is equal to \textit{physical capacity}. For example, on most microprocessors, L1 data cache's effective and physical capacity are identical, because it is not shared with other cores or instruction cache, and virtually mapped.
In contrast, a higher level cache for the same architecture might be shared among cores; contain the images of all those cores' L1 instruction caches or hold page tables, locked into L2 or L3 by hardware that walks the page table. Each of these effects might reduce the \textit{effective cache capacity} and modern commodity processors exhibit all three.

\textit{\textbf{\large Contribution}~~~}
This paper presents a set of tools that measure, efficiently and empirically, the \textit{effective capacity} and other parameters of the various levels in the data memory hierarchy, both cache and TLB;
that are portable across a variety of systems; 
that include a robust automatic analysis;
and that derive a full set of characteristics in a few seconds.
The resulting tools are inexpensive enough to use in a variety of contexts that may include install time, compile time or runtime adaption, or performance understanding tools.
Section~\ref{experiment} shows that our techniques produce results with the same accuracy as earlier work, while using a factor of~\textbf{10x} to~\textbf{250x} less time, which makes the tools inexpensive enough to be used in various contexts, especially lightweight runtime adaption.

\section{Related Work}\label{related-work}
Many authors describe systems
that attempt to characterize the memory
hierarchy~\cite{saavedra1995,mcvoy1996,dongarra2004,yotov2005,yotov2005a,sandoval2011},
but from our perspective, previous systems suffer from several specific flaws: (1) The prior tools are not easily portable to current machines~\cite{saavedra1995,mcvoy1996,dongarra2004}.
Some rely on system-specific features such as superpages or hardware performance counters to simplify the problems. Others were tested on older systems with shallow hierarchies; multi-level caches and physical-address tags create complications that they were not designed to handle. On contrast, our tools characterize various modern processors using only portable C code and \textsc{Posix} calls. 
(2) Some previous tools solve multiple parameters at once~\cite{yotov2005,yotov2005a}, which is not robust since if the code generates one wrong answer for one parameter, it inevitably causes the failure of all the other parameters. Noisy measurements and new hardware features, such as sharing or victim caches, can also cause these tests to produce inaccurate results. 
(3) Finally, the time cost of measuring a full set of characteristics by previous tools is very large, e.g. 2-8 minutes by Sandoval's tool~\cite{sandoval2011} (See Section~\ref{experiment}). 
At that cost, the set of reasonable applications for these measurements is limited. For these techniques to find practical use, the cost of measurement and analysis must be much lower.

The other related works have different focuses with ours. \textsc{P-Ray}~\cite{duchateau2008} and \textsc{Servet}~\cite{gonzalez2010} characterized
sharing and communication aspects of multi-core clusters. Taylor \textit{et al.}~\cite{taylor2010} extended memory characterization
techniques to AMD GPUs. Sussman \textit{et al.}~\cite{sussman2011}
arbitrated between different results produced from different benchmarks. Abel ~\cite{abel2012measurement} measured physical capacities of caches. \textsc{Servet} 3.0~\cite{gonzalez2013s} improved \textsc{Servet} \cite{gonzalez2010} by characterizing the network performance degradation. Casas \textit{et al.}~\cite{casas2014active,casas2016evaluation} quantified applications' utilization of the memory hierarchy.

\section{The Algorithms}\label{algorithms}
This section describes three tests we developed that measure the levels, capacity and
latency for cache and TLB, along with associativity
and linesize for L1 cache. All of the tests rely
on a standard C compiler and the \textsc{Posix} libraries for portability, and timings are taken based on \texttt{gettimeofday()}.
All the three algorithms rely on a data structure \textit{reference string}, implemented as
a circular chain of pointers, to create a specific pattern of memory references. The reference strings are different for each test to expose different aspects of the memory hierarchy, but their construction, running and timing are shared as presented below.

\vspace*{-.3cm}
\subsection{Reference Strings}\label{reference-string}
\begin{figure}[t]
\centering
\begin{minipage}{.5\textwidth}
\includegraphics[viewport= 60 465 480 720,clip,scale=0.4]{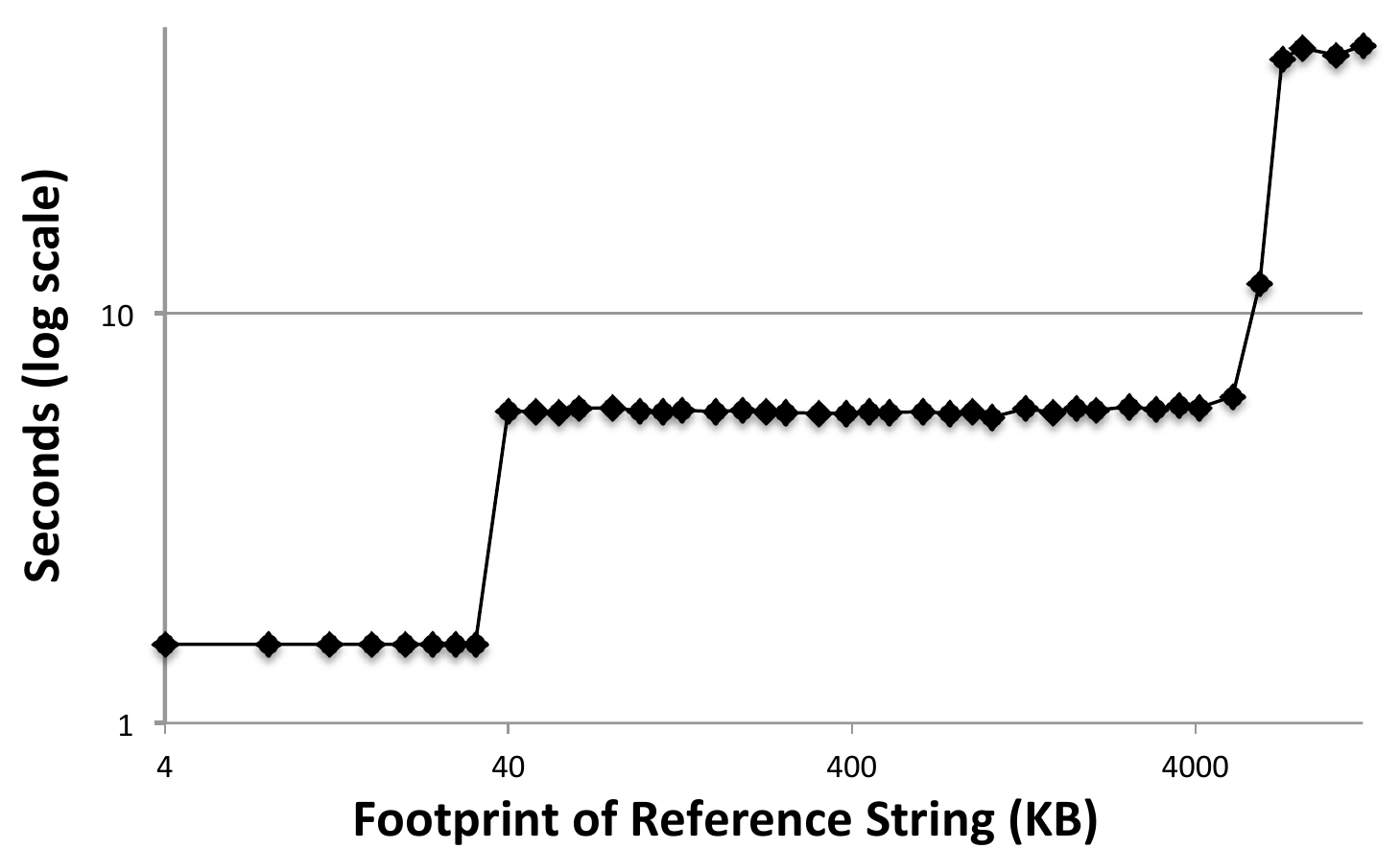}\\[-12pt]
\begin{picture}(0,0)
\put(70,34){\vector(-1,-1){10}}
\put(70,38){\small{}32\textsc{kb}}
\put(160,41){\vector(-1,1){10}}
\put(150,33){\small{}5\textsc{mb}}
\end{picture}
\caption{Example Response Curve} \label{example-response-curve}
\end{minipage}
\vspace*{.2cm}
\begin{minipage}{.45\textwidth}\noindent\rule{0.2in}{0pt}
\begin{algorithmic}
\State $loads \gets$ number of accesses
\State $start \gets$ timer()
\While {$loads-- > 0$}
	\State $p \gets *p$
\EndWhile
\State $finish \gets$ timer()
\State $elapsed \gets finish - start$
\end{algorithmic}
\vspace*{.5cm}
\caption{Running a Reference String} \label{running-a-rs}
\end{minipage}
\vspace*{-.6cm}
\end{figure}

 A reference string is simply a series of memory
references---in this paper, they are all load-operations---that the test uses to elicit a desired response from the memory system.
A reference string has a \textit{footprint}, the amount of contiguous virtual address
space that it covers. 
In general, the tests use a specific reference string, in a variety of footprints, to measure memory system response. 
By constructing, running different
footprints and recording the times spent, the test builds up a response curve. Fig.~\ref{example-response-curve} shows a typical response curve running on an Intel T9600.

\vspace*{.3cm}
\noindent\textbf{Running a Reference String:}
The microbenchmarks depend on the fact that we can produce an accurate measurement of the time required to run a reference string, and that we can amortize out compulsory start-up cache misses. To measure the running time for a reference string, the tool must instantiate the string and walk its references enough times to obtain an accurate timing.
Our tools instantiate the reference string as an array of pointers
whose size equals the footprint of the string.
Inside this array, the tools build a circular linked list of the
locations. (In C, we use an array of $void**$.)
The code to run the reference string is simple as shown in Fig.~\ref{running-a-rs}.
The actual implementation runs the loop enough times to
obtain stable results, where ``enough'' is scaled by the processor's measured speed.
We chose the iteration count by experimentation where the error rate is 1\% or less.

\vspace*{.3cm}
\noindent\textbf{Timing a Reference String:}
To time a reference string, we insert calls to the the $\textsc{Posix}$ \textit{gettimeofday} routine around the loop as a set of calipers.
We run multiple trials of each reference string
and record the minimum measured execution time because outside interference will only manifest itself in longer execution times. Finally, we convert the measured times into ``cycles'', where a cycle is defined as the time of an integer add operation. 
This conversion eliminates the fractional cycles
introduced by amortized compulsory misses and loop overhead.

The basic design and timing of reference strings are borrowed from Sandoval's work~\cite{sandoval2011}, since he already showed that these techniques produce accurate
timing results across a broad variety of processor
architectures and models. Our contribution lies in finding significantly more efficient ways to manipulate the reference strings to detect cache parameters. The L1 cache test provides a good
example of how changing the use of the string can
produce equivalent results with an order of magnitude
less cost (see Section~\ref{experimental-validation}
for cost and accuracy results).
Also, we significantly reduce the measurement time of data while keeping the same accuracy. We repeat each test until we have not seen the minimum time change in the last 25 runs. (The value of 25 was selected via a parameter sweep from 1 to 100. At~25, the error rate for the multi-cache test is~1\% or less.) 

\subsection{L1 Cache} \label{gap-test} \label{l1-cache-test}
Because the L1 data-cache linesize can be used to reduce spatial locality in the multi-level cache test and the TLB test, we use an efficient, specialized test to discover the L1 cache parameters.
Two properties of L1 caches make the test easy to derive and analyze: the lack of sharing at L1 and the use of virtual-address tags.
The L1 test relies directly on hardware effects caused by the combination of capacity and associativity.
We denote the L1 reference string as a tuple G(n,k,o), where n is the
number of locations to access, k is the number of bytes between those
locations (the ``gap''), and o is an offset added to the start of the last location in the set. The reference string G(n,k,0) generates the following locations:

\begin{center}
\vspace*{-.2cm}
\begin{picture}(150,22)(0,9)
\put(  0,20){\framebox(10,10){}}
\put(  0,23){\makebox[10pt][c]{\iname{s}{1}}}
\put( 40,20){\framebox(10,10){}}
\put( 40,23){\makebox[10pt][c]{\iname{s}{2}}}
\put( 80,20){\framebox(10,10){}}
\put( 80,23){\makebox[10pt][c]{\iname{s}{3}}}
\put(115,23){\makebox[0pt][c]{\ldots}}
\put(140,20){\framebox(10,10){}}
\put(140,23){\makebox[10pt][c]{\iname{s}{n}}}
\put(  0,15){\line(1,0){150}}
\put(  0,15){\line(0,1){3}}
\put(  0, 9){\makebox[0pt][l]{$_0$}}
\put( 40,15){\line(0,1){3}}
\put( 40, 9){\makebox[0pt][l]{$_k$}}
\put( 80,15){\line(0,1){3}}
\put( 80, 9){\makebox[0pt][l]{$_{2k}$}}
\put(140, 9){\makebox[0pt][l]{$_{(n-1)k}$}}
\put(140,15){\line(0,1){3}}
\end{picture}
\end{center}

\noindent
And \textit{G(n,k,\textbf{4})} would move the \textit{n}$^{th}$
location out another four bytes.

Both X-Ray~\cite{yotov2005} and Sandoval's~\cite{sandoval2011} ``gap'' test use a similar reference string.
However, they require to iterate \textit{k} across the full
range of cache sizes, and \textit{n} from 1 to the actual associativity plus one.
Our algorithm orders the tests in a different way that radically
reduces the number of combinations of \textit{n} and \textit{k} that it must try.
It starts with the maximum value of
associativity, \textbf{MaxAssoc}, say 16, and sweeps over the gap size to find the first value that causes significant misses, as shown in the first loop in Fig.~\ref{pseudo-code-for-the-l1-cache-test}.

\begin{figure}[t]
\centering
\makebox[\textwidth]{\makebox[1.1\textwidth]{
\begin{minipage}{.55\textwidth}
$baseline \gets$ time for G(2,LB/2,0)\\
\noindent\textbf{1. for} $k \gets$ LB/MaxAssoc to UB/MaxAssoc \\
\hspace*{.7cm}$t \gets $ time for G(MaxAssoc+1,k,0) \\
\hspace*{.7cm}\textbf{if} {$ t > baseline $}\\
		\hspace*{1.2cm} $L1Size = k * MaxAssoc$\\
		\hspace*{1.2cm} $break$ \\	  	
\end{minipage}
\hspace*{.2cm}
\begin{minipage}{.45\textwidth}
\noindent\textbf{2. for} {$n \gets MaxAssoc; n \geq 1; n \gets n/2 $}\\
	\hspace*{.7cm} $t \gets$ time for G(n+1,L1Size/n,0) \\
	\hspace*{.7cm} \textbf{if} {$ t \leq baseline $}\\
		\hspace*{1.2cm} $L1Assoc = n * 2$\\
		\hspace*{1.2cm} $break$ \\
\end{minipage}}}
\subfloat{
\begin{minipage}{1.1\textwidth}
\noindent\textbf{3. for} offset $\gets$ 1 to pagesize \\
	\hspace*{.7cm} $t \gets$ time for G(L1Assoc+1,L1Size/L1Assoc,offset)\\
	\hspace*{.7cm} \textbf{if} $ t \leq baseline $\\
		\hspace*{1.2cm} L1LineSize = offset\\
		\hspace*{1.2cm} $break$ \\
\end{minipage}}
\vspace*{-0.5cm}
\caption{Pseudo Code for the L1 Cache Test}\label{pseudo-code-for-the-l1-cache-test}
\vspace*{-0.5cm}
\end{figure}
\begin{figure}[t]
\makebox[\textwidth]{\makebox[1.2\textwidth]{
\begin{minipage}{.5\textwidth}
\centering
\includegraphics[keepaspectratio=true,scale=0.35]
                     {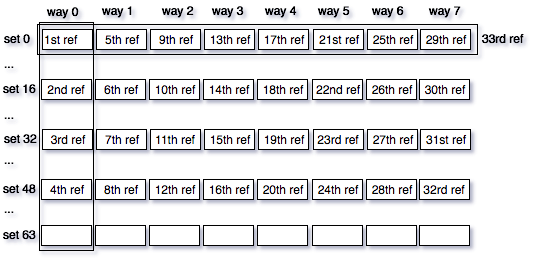}
\caption{Running G(33,1KB,0) on a 32KB, 8-Way, Set-Associative Cache}\label{gap-example}
\end{minipage}
\hspace*{.2cm}
\begin{minipage}{.5\textwidth}
\centering
\includegraphics[keepaspectratio=true,scale=0.35]
                     {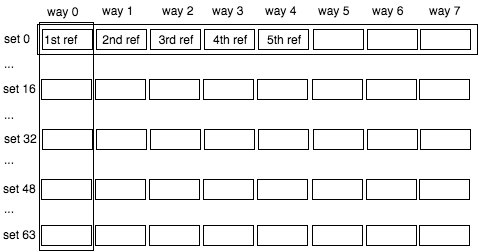}
\caption{Running G(5,8KB,0) on a 32KB, 8-Way, Set-Associative Cache}
\label{gap-find-assoc}
\end{minipage}}}
\vspace*{-.5cm}
\end{figure}

It sweeps the value of \textit{k} from
 LB\,/\,MaxAssoc to UB\,/\,MaxAssoc, where  \textbf{LB} and  \textbf{UB} are the lower and upper bounds of the testing cache. For L1 cache, we choose 1\textsc{KB} and 4\textsc{MB} respectively. 
With $\textit{n} = \textrm{MaxAssoc}+1$ and $\textit{o} = 0$ ,
the last and the first location will always map to the same
cache set location.
None of them will miss until $k*MaxAssoc$ reaches the L1 cache size.
Fig.~\ref{gap-example} shows how the string G(33,\,1\textsc{kb},\,0)
maps into a 32\textsc{kb}, 8-way, set-associative  cache.
With these cache parameters, each way holds 4\textsc{kb} and all
references in G(33,\,1\textsc{kb},\,0) map into sets 0, 16, 32, and 48.
The string completely fills those four cache sets, and overflows
set~0. Thus, the 33$^{rd}$ reference will always miss in the L1 cache.

The first loop will record uniform (cycle-rounded) times for
all iterations, which is called the \textbf{baseline} time since all the references hit in L1 cache so far.  Until it reaches G(33,\,1\textsc{kb},\,0), at which
time it will record a larger time due to the miss on the 33$^{rd}$
element, the larger time causes it to record the cache size and
exit the first loop.

The second loop in Fig.~\ref{pseudo-code-for-the-l1-cache-test}
finds the associativity by looking at larger gaps (\textit{k}) and
smaller associativities (\textit{n}).
It sweeps over associativity from MaxAssoc\,+\,1 to 2, 
decreasing $n$ by a factor of 2 at each iteration.
In a set-associative cache, the last location in all the reference strings
will continue to miss in L1 cache until n is less than the actual associativity.
In the example, a 32KB, 8 way L1 cache, that occurs when n is L1Assoc\,/\,2,
namely 4, as shown in Fig.~\ref{gap-find-assoc}.
At this point, the locations in the reference string all map to the
same set and, because there are more ways in the set than references now,
the last reference will hit in cache and the time will again match the baseline time.

At this point, the algorithm knows the L1 size and associativity.
So the third loop runs a parameter sweep on the value of \textit{o}, from 1 to pagesize (in words).
When \textit{o} reaches the L1 linesize, the final reference in the string maps to the next set and the runtime for the reference string drops back to the original baseline---the value when all references hit in cache.

\subsection{Multilevel Caches}\label{multi-cache}\label{multi-level-cache-test}

The L1 cache test relies on the fact that it can precisely detect the actual hardware boundary of the cache.
We cannot apply the same test to higher level caches for several reasons:
higher level caches tend to be shared, either between instruction and data cache, or between cores, or both;
higher level caches tend to use physical-address tags rather than virtual ones;
operating systems tend to lock page table entries into one of the higher level caches. 
Each of these factors, individually, can cause the
L1 cache test to fail.
It works on L1 precisely because L1 data caches are
core-private and virtually mapped, without outside interference
or page tables.

Our multi-level cache test avoids the
weaknesses that the L1 cache test exhibits for upper level caches by
detecting cache capacity in isolation from associativity.
It uses a reference string designed to expose changes in cache response while  
isolating those effects from any TLB response.
It reuses the infrastructure developed for the L1 cache test
to run and time the cache reference string.

The multi-level cache test reference string $C(k)$ is constructed from a footprint \textit{k}, the OS pagesize obtained from the
\textit{Posix} \texttt{sysconf()}, and the L1 cache linesize.\footnote{In
practice, the L1 linesize is used to accentuate the system response by decreasing spatial locality. Any value greater than $sizeof(void*)$ works, but a value greater than or equal to linesize works better.}
Given these values, the string generator builds an array of pointers that
spans \textit{k} bytes of memory.
The generator constructs an index set, the column set, that covers one page
and accesses one pointer in each L1 cache line on the page.
It constructs another index set, the row set, that contains the starting
address of each page in the array.
Fig.~\ref{cache-reference} shows the cache reference string without
randomization; in practice, we randomize the order within each row and the order of the the rows to eliminate the impact of hardware prefetching schemes.
Sweeping through the page in its randomized order before moving to another page, minimizes the impact of TLB misses on large footprint reference strings.

To measure cache capacity, the test could use this reference string in a simple parameter sweep, for the range from  \textbf{LB} to  \textbf{UB}. As we mentioned,  \textbf{LB} and  \textbf{UB} are the lower and upper bounds of the testing cache and we choose 1\textsc{KB} and 32\textsc{MB} respectively for the whole multi-level cache.

\vspace{6pt}
\noindent\rule{0.2in}{0pt}\begin{minipage}{3in}
\textbf{for} {$k \gets$ Range(LB,UB)}\\
	\hspace*{.5cm}$t_{k} \gets$ time for $C(k)$ reference string
\end{minipage}
\vspace{6pt}

The sweep produces a series of values, $t_{k}$, that form a piecewise
linear function describing the processor's cache response.  Recall that Fig.~\ref{example-response-curve} shows the curve from the multi-level cache test on an Intel T9600. 
The T9600 featured a 32\textsc{kb} L1 cache and a 6\textsc{mb} L2 cache, but notice the sharp rise at 32\textsc{kb} and the softer rise that begins at 5\textsc{mb}. It's the effect of \textit{effective capacity}.

The implementation of the algorithm, of course, is more complex.
Section~\ref{reference-string} described how to run and time a single reference string. Besides that, the pseudo code given inline above abstracts the choice of sample points into the notation Range(LB,UB).
Instead of sampling the whole space uniformly, in both the multi-level cache and the TLB test, we actually only space the points uniformly below 4\textsc{kb} (we test 1, 2, 3, and~4\textsc{kb}).
Above 4\textsc{kb}, we test each power of two, along with three
points uniformly-spaced in between since sampling fewer points has a direct effect on running time.
The pseudo code also abstracts the fact that the test actually makes repeated sweeps over the range from LB to UB. At each size, it constructs, runs, and times a reference string, updating the minimal time for that size, if necessary, and tracking
the number of trials since the time was last lowered. Sweeping in this way distributes outside interference, say from an OS daemon or another process, across sizes, rather than concentrating it in a small number of sizes. Recreating the reference string at each size and trial allows the algorithm to sample different virtual-to-physical page mappings.
\begin{figure}[t]
\makebox[\textwidth]{\makebox[1.1\textwidth]{
\begin{minipage}{.5\textwidth}
\vspace*{-6pt}
\begin{center}
\begin{picture}(180,110)(-20,0)\thinlines
\multiput(0,0)(0,10){8}{\color[cmyk]{0,0,0,0.7}
  \multiput(0,0)(20,0){8}{\framebox(20,10){}}
}
\multiput(9,76)(0,-10){8}{
  \multiput(0,0)(20,0){7}{\qbezier(0,0)(10,5)(16,1)}
}
\multiput(24,78)(0,-10){8}{
  \multiput(0,0)(20,0){7}{\vector(1,-1){3}}
}
\multiput(152,76)(0,-10){7}{\qbezier(0,0)(0,-3,)(-10,-3)}
\multiput(142,73)(0,-10){7}{\line(-1,0){120}}
\multiput( 22,73)(0,-10){7}{\qbezier(0,0)(-14,0)(-16,-5)}
\multiput(  7,70)(0,-10){7}{\vector(-1,-2){3}}
\qbezier(150, 3)(158,3)(158,15)
\put(158,15){\line(0,1){60}}
\qbezier(158,75)(158,83)(150,83)
\put(150,83){\line(-1,0){128}}
\qbezier(22,83)(8,83)(6,78)
\put( 7,80){\vector(-1,-2){3}}
\put(00,87){\line(0,1){10}}
\put(20,87){\line(0,1){10}}
\put( 4,89){\vector(1,0){16}}
\put(16,89){\vector(-1,0){16}}
\put( 0,92){\makebox[19pt][c]{\scriptsize\itshape 1\,line}}
\put( 0, 101){\line(0,1){7}}
\put(160,101){\line(0,1){7}}
\put(67, 104){\vector(-1,0){67}}
\put(93,104){\vector(1,0){67}}
\put(68,102){\makebox[24pt][c]{\scriptsize\itshape 1\,page}}
\put(-40,40){\parbox{20pt}{
     \begin{center}\scriptsize\itshape k/pagesize\\[-2pt]rows\end{center}}}
\put(-30,35){\vector(0,-1){35}}
\put(-34,0){\line(1,0){8}}
\put(-30,50){\vector(0,1){30}}
\put(-34,80){\line(1,0){8}}
\end{picture}
\end{center}
\vspace{-12pt}
\caption{Cache Test Reference String}
\label{cache-reference}
\vspace*{-6pt}
\end{minipage}
\begin{minipage}{.5\textwidth}
\begin{center}
\begin{picture}(180,110)(-20,0)\thinlines
\multiput(0,0)(0,10){8}{\color[cmyk]{0,0,0,0.7}
  \multiput(0,0)(20,0){8}{\framebox(20,10){}}
}
\multiput(10,75)(20,-10){7}{\qbezier(0,-2)(0,-10)(17,-10)}
\multiput(27,65)(20,-10){7}{\vector(1,0){3}}
\qbezier(152,2)(152,76)(14,76)
\put(14,76){\vector(-1,0){3}}
\put(00,87){\line(0,1){10}}
\put(20,87){\line(0,1){10}}
\put( 4,89){\vector(1,0){16}}
\put(16,89){\vector(-1,0){16}}
\put( 0,92){\makebox[19pt][c]{\scriptsize\itshape 1\,line}}
\put( 0, 101){\line(0,1){7}}
\put(160,101){\line(0,1){7}}
\put(67, 104){\vector(-1,0){67}}
\put(93,104){\vector(1,0){67}}
\put(68,102){\makebox[24pt][c]{\scriptsize\itshape 1\,page}}
\end{picture}
\end{center}
\vspace{-12pt}
\caption{TLB Test Reference String}
\label{tlb-test-reference-string}
\end{minipage}}}
\vspace*{-.8cm}
\end{figure}

\vspace*{.3cm}
\noindent\textbf{Knocking Out and Reviving Neighbors:}
Most of the time spent in the multi-level cache test is incurred by running reference strings. The discipline of running each size until its minimum time is ``stable''---defined as not changing in the last 25 runs, means that the test runs enough reference strings. (As we mentioned, the value of 25 was selected via a parameter sweep from 1 to 100 and at~25, the error rate for the multi-cache test is~1\% or less.) 

In Fig.~\ref{example-response-curve}, points that fall in the middle of a level of the memory hierarchy have, as might be expected, similar heights in the graph, indicating similar average latencies.
Examining the sweep-by-sweep results, we realized that values in those plateaus quickly reach their minimum values. This is another kind of ``stability'' of data.
To capitalize on this effect, we added a mechanism to knock values out of the testing range when they agree with the values at neighboring sample sizes.
As shown in Fig.~\ref{cache-test}, after every sweep of the reference
strings, the knockout phase examines every sample size $k$ in the Range(LB, UB).
If $t_{k}$, the time of running $C(k)$ string, equals both $t_{k-1}$
and $t_{k+1}$, then it cautiously asserts $k$ is a redundant sample size
and can be knocked out. (It sets the counter
for that size that tracks iterations since a new minimum to zero.)
In the next sweep of Range(LB,\,UB), these sample sizes will be omitted.

The knock-out mechanism can eliminate a sample size too soon, e.g. a sample size $k$ and its neighbors have the same inaccurate value.
When this occurs, the knockout test may eliminate out $k$ prematurely.
To cope with this situation, we added a revival phase.
When any point reaches a new minimum, if it has a neighbor that was previously knocked
out, it revives that neighbor so that it is again measured in the next sweep of Range(LB,\,UB).

The knockout-revival optimization significantly reduced the cost of the multi-level cache test, from minutes in earlier work to seconds, as details shown in Section~\ref{experiment}.


\subsection{TLB Test} \label{tlb-only-test}
The TLB test closely resembles the multi-level cache test, except that it
uses a reference string that isolates TLB behavior from cache misses.
It utilizes the same infrastructure to run reference strings and incorporates the
same discipline for sweeping a range of TLB sizes to produce a response curve.
It benefits significantly from the knockout-revival mechanism.

The TLB reference string, $T(n,k)$, accesses $n$ pointers in each
page of an array with a footprint of $k$ bytes.
To construct $T(1,k)$, the generator builds a column index set and
a row index set as in the multi-level cache test.
It shuffles both sets.
To generate the permutation, it iterates over the row set choosing pages.
It chooses a single line within the page by using successive lines from
the column set, wrapping around in a modular fashion if necessary.
The result is a string that accesses one line per page, and spreads the lines
over the associative sets in the lower level caches.
The effect is to maximize the page footprint, while minimizing the
cache footprint.
Fig.~\ref{tlb-test-reference-string} shows $T(1, k)$ without randomization.
For $n>1$, the generator uses $n$ lines per page, with a variable offset within the page to distribute the accesses across different sets in the caches and minimize associativity conflicts. The generator randomizes the full set of references, to avoid the effects of a prefetcher and to successive accesses to the same page.


The multi-level cache test hides the impact of TLB misses by amortizing those misses over many accesses.
Unfortunately, the TLB test cannot completely hide the impact of cache misses because any action that amortizes cache misses also partially amortizes TLB misses.
When the TLB line crosses a cache boundary, the rise in measured time is indistinguishable from the response to a TLB boundary.
However, we could rule out false positives by running $T(2,k)$ reference string and following the rule that if $T(1,k)$ shows a TLB response at $x$ pages, then $T(2,k)$ should show a TLB response at $x$ pages too if x pages is indeed a boundary of TLB.
Because $T(2,k)$ uses twice as many lines at $x$ pages as $T(1,k)$,
if it's a false positive response caused by the cache boundary in $T(1,k)$, it will appear at a smaller size in $T(2,k)$.

\begin{figure}[t]
\centering
\makebox[\textwidth]{\makebox[1.1\textwidth]{
\begin{minipage}{.6\textwidth}
\textbf{while} {not all $t_{k}$ are stable} \textbf{do}\\
\hspace*{.5cm}\textbf{for} {$k \gets$ Range(LB,UB)}\\
	 \hspace*{1cm}$t_{k} \gets$ time for $C(k)$ reference string\\
	 \hspace*{1cm}\textbf{if} {$t_{k}$ is a new minimum \&\& $k$'s neighbors\\
     \hspace*{1cm}have been knocked out}\\
			\hspace*{1.5cm}revive $k$'s neighbors to Range(LB,UB)
\hspace*{.5cm}\textbf{for} {$k \gets$ Range(LB,UB)}\\
	\hspace*{1cm}\textbf{if} {$t_{k} == t_{k's\ neighbors}$ }\\
		\hspace*{1.5cm}knock out $k$ from Range(LB, UB)\\
\vspace*{-10pt}
\caption{Pseudo Code for Multi-Level Cache Test} \label{cache-test}
\end{minipage}
\hspace*{.2cm}
\begin{minipage}{.5\textwidth}
\textbf{for} {$k \gets$ LB to UB}\\
	\hspace*{.5cm}$t_{1,k} \gets$ time for $T(1,k)$\\
	\hspace*{.5cm}\textbf{if} {there's a jump from $t_{1,k-1}$ to $t_{1, k}$}\\
		\hspace*{1cm}\textbf{for} {$n \gets$ 2, 3, 4}\\
			\hspace*{1.5cm}$t_{n,k-1} \gets$ time for T(n,k-1)\\
			\hspace*{1.5cm}$t_{n,k} \gets$ time for T(n,k)\\ 
		\hspace*{1cm}\textbf{if} {there's a jump from $t_{n,k-1}$ to $t_{n, k}$\\
        \hspace*{1cm}when n=2,3,4}\\
			\hspace*{1.5cm}report a TLB size\\
\vspace*{-10pt}
\caption{Pseudo Code for the TLB Test} \label{tlb-test}
\end{minipage}}}
\vspace*{-.5cm}
\end{figure}

Still, a worst-case choice of cache and TLB sizes can fool this test.
If $T(1,k)$ maps into $m$ cache lines at $x$ pages, and $T(2,k)$ maps
into $2*m$ cache lines at $x$ pages, and the processor has cache boundaries at  $m$ and $2*m$ lines, both reference strings will discover a suspect
point at $x$ pages and the current analysis will report a TLB boundary at $x$ pages even if it's not. Using more tests, e.g., $T(3,k)$ and $T(4,k)$, could eliminate these false positive points.

Sandoval's test~\cite{sandoval2011} ran the higher line-count TLB strings, $T(2,k)$, $T(3,k)$, and $T(4,k)$ exhaustively.
We observe that the values for those higher line-count tests are only of interest at points where the code observes a transition in the response curve.
Thus, our TLB test runs a series of $T(1,k)$ strings for $k$ from $LB$ to $UB$.
From this data, it identifies potential transitions in the TLB response graph, called ``suspect" points. Then it examines the responses of the higher line-count tests at the suspect point and its immediate neighbors as shown in Fig.~\ref{tlb-test}.
If the test detects a rise in the $T(1,k)$ response at \textit{x} pages, but that response is not confirmed by one or more of $T(2,k)$, $T(3,k)$, or $T(4,k)$, then it reports \textit{x} as a false positive.
If all of $T(1,k)$, $T(2,k)$, $T(3,k)$, $T(4,k)$ show a rise at $x$ pages, it reports that transition as a TLB boundary.
This technique eliminates almost all false positive results in practice since the situation that all $m$, $2m$, $3m$, $4m$ cache lines are cache boundaries is extremely unlikely.

Running the higher line-count tests as on-demand confirmatory tests, rather than exhaustively, significantly reduces the number of reference strings run and, thus, the overall time for the TLB test.(See time cost comparison in Section~\ref{experiment}.)


\section{Experimental Validation}\label{experiment}\label{experimental-validation}
\begin{table*}[t]
  \centering
  \def\arraystretch{1.25}
\begin{tabular}{|c|c|c|c|c|c|} \hline
Processor&Capacity(\textsc{kb})&LineSize (B)&Associativity&Latency Cycle&Cost Secs\\
\hline
Intel Core i3 & 32 & 64 &8&3 &0.49\\ \hline
Intel Core i7 & 32 & 64 &8&5 &0.56\\ \hline
Intel Xeon E5-2640 & 32 & 64 &8&4 &0.49\\ \hline 
Intel Xeon E5420 & 32 & 64 &8&4 &0.54\\ \hline
Intel Xeon X3220 & 32 & 64 &8&3 &0.51\\ \hline
Intel Xeon E7330 & 32 & 64 &8&3 &0.52\\ \hline
Intel Core2 T7200 & 32 & 64 &8&3 &0.55\\ \hline
Intel Pentium 4 & 8 & 64 &4&4 &1.71\\ \hline
ARM Cortex A9&  32& 32 &7 & 4& 7.98\\ \hline
\end{tabular}
\vspace*{0.1cm}
\caption{L1 Cache Results}
\label{l1cache-results}
\vspace*{-.8cm}
\end{table*}
To validate our techniques, we ran them on a collection
of systems that range from commodity X86 processors through an ARM Cortex~A9.
All of these systems run some flavor of Unix and support
the \textsc{Posix} interfaces for our tools.

Table~\ref{l1cache-results} shows the results of the L1 cache test: the capacity, line size, associativity, and latency,
along with the total time cost of the measurements.
On most machines, the tests only required roughly half a second to detect all the L1 cache parameters, except for the ARM machine and the Pentium 4. 
The ARM timings are much slower in all three tests because it is a Solid Run Cubox-i4Pro with a 1.2\textsc{ghz} Freescale iMX6-Quad processor. It runs Debian Unix from a commercial SD card in lieu of a disk. Thus, it has a different OS, different compiler base, and different hardware setup than the other systems, which are all off-the-shelf desktops, servers, or laptops.
Thus all of the timings from the ARM system are proportionately slower than the other systems.
The Intel Pentium 4 system is relatively higher than the other chips, despite its relatively fast clock speed of 3.2\textsc{ghz}.
Two factors explain this seemingly slow measurement time.
The latency of the Pentium~4's last level cache is slow relative to most modern systems. Thus, it runs samples that hit in the cache more slowly than the other modern systems. In addition, its small cache size (384\textsc{kb}) means that a larger number
of samples miss in the last level cache (and run slowly in main memory) than the other tested systems.

\begin{table*}[t]
\centering
\def\arraystretch{1.5}
\begin{tabular}{|c|c|c|c|c|c|} \hline
Processor&Level &Effective Cap.(\textsc{kb})&Physical Cap.(\textsc{kb})&Latency Cycle&Cost(Secs)\\
\hline
Intel Core i3 & \shortstack{\\1\\ 2\\3} & \shortstack{ 32\\256\\\textbf{2048}}& \shortstack{ 32\\256\\3072} &\shortstack{ 3\\13\\30} &2.05\\ \hline
Intel Core i7 & \shortstack{\\1\\ 2\\3} & \shortstack{ 32\\256\\\textbf{3072}}& \shortstack{ 32\\256\\4096} &\shortstack{ 5\\13\\36} &3.70\\ \hline
Intel Xeon E5-2640 & \shortstack{\\1\\ 2\\3} & \shortstack{ 32\\256\\\textbf{14336}}& \shortstack{ 32\\256\\15360} &\shortstack{ 4\\13\\42} &4.49\\ \hline
Intel Xeon E5420 & \shortstack{\\1\\ 2} & \shortstack{ 32\\\textbf{4096}}& \shortstack{ 32\\6144} &\shortstack{ 4\\16} &4.68\\ \hline
Intel Xeon X3220 & \shortstack{\\1\\ 2} & \shortstack{ 32\\\textbf{3072}}& \shortstack{ 32\\4096} &\shortstack{ 3\\15} &3.92\\ \hline
Intel Xeon E7330 & \shortstack{\\1\\ 2} & \shortstack{ 32\\\textbf{1792}}& \shortstack{ 32\\3072} &\shortstack{ 3\\14} &6.87\\ \hline
Intel Core2 T7200 & \shortstack{\\1\\ 2} & \shortstack{ 32\\4096}& \shortstack{ 32\\4096}&\shortstack{ 3\\15} &5.66\\ \hline
Intel Pentium 4 & \shortstack{\\1\\ 2} & \shortstack{ 8\\\textbf{384}} & \shortstack{ 8\\512}&\shortstack{ 4\\39} &8.03\\ \hline
ARM Cortex A9& \shortstack{\\1\\ 2} & \shortstack{ 32\\1024} & \shortstack{ 32\\1024}&\shortstack{ 4\\11} & 54.57\\ \hline
\end{tabular}
\vspace*{0.1cm}
\caption{Multilevel Caches Results}
\label{cache-results}
\vspace*{-.8cm}
\end{table*}

The multi-level cache and TLB tests produce noisy data that approximates the piecewise linear step functions that describe the processor's response. We developed an  automatic, conservative and robust analysis tool which uses a multi-step process to derive consistent and accurate capacities from that data. \footnote{
The details are omitted due to space limit. Please contact the authors if interested.}
The analysis derive for both cache and TLB, the number of levels and the transition point between each pair of levels (\textit{i.e.}, the effective capacity of each level). 

Table~\ref{cache-results} shows the measured parameters from the multi-level cache test: the number of cache levels,
effective capacity, latency, and total time required for the
measurement.
In addition, the table shows the actual physical capacities for comparison against the effective capacities measured by the tests.
The tests are precise for lower-level caches, but typically underestimate the last level of cache---that is, their effective capacities are smaller than the physical cache sizes for the last level of cache.
As discussed earlier, if the last level of cache is shared by multiple cores, or if the OS locks the page table into that  cache, we would expect the effective capacity to be smaller than the physical capacity.
The time costs of the multi-level cache test are all few seconds except for the ARM machine because of the same reasons we explained above.

Table~\ref{tlb-results} shows the results of TLB test:
the number of levels, effective capacity for
each level (\textit{entries $\times$ pagesize}), and the time cost of the measurement.
From Table~\ref{tlb-results} we see that, on most systems, the TLB test ran in less than~1 second.
As with the multi-level cache test, the Pentium~4 is slower than the newer systems. The same factors come into play. It has a small, one-level TLB with a capacity of 256\textsc{kb}.
The test runs footprints up to 8\textsc{mb}, so the Pentium~4 generates many more TLB misses than are seen on machines with larger TLB capacities.

\begin{table}[t]
\centering
\begin{tabular}{|c|c|c|c|} \hline
Processor& Level &Capacity (\textsc{kb})&Cost (Secs)\\
\hline
Intel Core i3 & \shortstack{\\1\\ 2} & \shortstack{ 256\\2048} &0.14\\ \hline
Intel Core i7 & \shortstack{\\1\\ 2} & \shortstack{ 256\\4096}  &0.84\\ \hline
Intel Xeon E5-2640 & \shortstack{\\1\\2} &\shortstack{ 256\\2048} &0.15\\ \hline
Intel Xeon E5420 & \shortstack{\\1\\2} & \shortstack{64\\1024} &0.20\\ \hline
Intel Xeon X3220 & \shortstack{\\1\\2} & \shortstack{64\\1024} &0.18\\ \hline
Intel Xeon E7330 & \shortstack{\\1\\2} & \shortstack{64\\1024} &0.20\\ \hline
Intel Core2 T7200 & \shortstack{\\1\\ 2} & \shortstack{ 64\\1024} &1.28\\ \hline
Intel Pentium 4 \rule{0pt}{12pt}& \shortstack{1} & \shortstack{256} &3.09\\ \hline
ARM Cortex A9& \shortstack{\\1\\2} & \shortstack{128\\512} &8.91\\ \hline

\end{tabular}
\vspace*{0.1cm}
\caption{TLB Results}
\label{tlb-results}
\vspace*{-.8cm}
\end{table}

The reason why we didn't measure the associativity and linesize for multi-level caches and TLB is that caches above L1 tend to use physical-address tags rather than
virtual-address tags, which complicates the measurements.
The tools generate reference string that are contiguous in virtual address space;
the distance relationships between pages in virtual space are not guaranteed to hold
in the physical address space. (Distances within a page hold in both spaces.)

In a sense, the distinct runs of the reference string form distinct
samples of the virtual-to-physical address mapping.
(Each time a specific footprint is tested, a new reference string is allocated
and built.) Thus, any given run at any reasonably large size can
show an unexpectedly large time if the virtual-to-physical
mapping introduces an associativity problem in the physical address space
that does not occur in the virtual address space.

The same effect makes it difficult to determine associativity at the upper
level caches. Use of physical addresses makes it impossible to create,
reliably, repeatedly, and port\-ably, the relationships required to measure
associativity. 
In addition, associativity measurement requires the ability to detect the
\textbf{actual}, rather than \textbf{effective}, capacity. 

The goal of this work was to produce efficient tests. We compare the time cost of our tool and Sandoval's~\cite{sandoval2011} as shown in Table~\ref{comparison}. (Other prior tools either cannot run on modern processors or produce wrong answers every now and then.) 
The savings in measurement time are striking.
On the Intel processors, the reformulated L1 cache test is \textbf{20} to \textbf{70} times faster;
the multi-level cache test is \textbf{15} to \textbf{40} times faster; and 
the TLB test is \textbf{60} to \textbf{250} times faster.
The ARM Cortex A9 again shows distinctly different timing results:
the L1 cache test is~2 times faster, the multi-level cache test is about~8.4 times
faster, and the TLB test is about~4.7 times faster.

\begin{table*}[t]
\centering
\begin{tabular}{|c|c|c|c|c|} \hline
Processor&Tools &L1 Test Cost&Multilevel Test Cost&TLB Test Cost\\
\hline
Intel Core i3 & \shortstack{\\\\Our Tool\\Sandoval's tool} & \shortstack{ 0.49\\27.02} & \shortstack{ 2.05\\58.16}& \shortstack{ 0.14\\36.81}\\ \hline
Intel Core i7 & \shortstack{\\\\Our Tool\\ Sandoval's tool}  & \shortstack{ 0.56\\34.75}  & \shortstack{ 3.70\\92.35}& \shortstack{ 0.84\\94.97}\\ \hline
Intel Xeon E5-2640 & \shortstack{\\\\Our Tool\\ Sandoval's tool}  &\shortstack{ 0.49\\33.33} & \shortstack{ 4.49\\65.42}& \shortstack{ 0.15\\38.79}\\ \hline
Intel Xeon E5420 & \shortstack{\\\\Our Tool\\ Sandoval's tool}  & \shortstack{0.54\\28.86} & \shortstack{ 4.68\\150.43}& \shortstack{ 0.20\\55.56}\\ \hline
Intel Xeon X3220 & \shortstack{\\\\Our Tool\\ Sandoval's tool}  & \shortstack{0.51\\28.89} & \shortstack{ 3.92\\121.54}& \shortstack{ 0.18\\54.17}\\ \hline
Intel Xeon E7330 & \shortstack{\\\\Our Tool\\ Sandoval's tool}  & \shortstack{0.52\\35.77} & \shortstack{ 6.87\\228.13}& \shortstack{ 0.20\\53.24}\\ \hline
Intel Core2 T7200 & \shortstack{\\\\Our Tool\\ Sandoval's tool}  & \shortstack{ 0.55\\34.86} & \shortstack{ 5.66\\200.82}& \shortstack{ 1.28\\166.19}\\ \hline
Intel Pentium 4 & \shortstack{\\\\Our Tool\\ Sandoval's tool}  & \shortstack{1.71\\40.45} & \shortstack{ 8.03\\227.57}& \shortstack{ 3.09\\194.37}\\ \hline
ARM Cortex A9 & \shortstack{\\\\Our Tool\\ Sandoval's tool}
     & \shortstack{7.98\\16.76}
     & \shortstack{54.57\\458.55}& \shortstack{8.91\\42.03}\\ \hline

\end{tabular}
\vspace*{0.1cm}
\caption{Our Tool VS. Sandoval's Tool}
\label{comparison}
\vspace*{-.8cm}
\end{table*}

\section{Conclusions}\label{conclusion}
This paper has presented techniques that efficiently measure the \textit{\textbf{effective
capacities}} and other performance-critical parameters of a processor's cache and TLB hierarchy.
The tools are portable; they rely on a C compiler and the \textsc{Posix~OS}
interfaces.
The tools are efficient; they take at most a few seconds to discover
effective cache and TLB sizes.
This kind of data has application in code optimization, runtime adaptation,
and performance understanding.

This work lays the foundation for two kinds of future work:
(1) measurement of more complex parameters, such as discovering the
sharing relationships among hardware resources, or measuring the
presence and capacities of features such as victim caches and
streaming buffers; and
(2) techniques for lightweight runtime adaptation, either with
compiled code that relies on runtime-provision of hardware parameters
or with lightweight mechanisms for runtime selection from pre-compiled
alternative code sequences.


\begin{thebibliography}{1}
\bibitem{saavedra1995}
Saavedra, R.H. and Smith, A.J., 1995. Measuring cache and TLB performance and their effect on benchmark runtimes. IEEE Transactions on Computers, 44(10), pp.1223-1235.

\bibitem{mcvoy1996}
McVoy, L.W. and Staelin, C., 1996. lmbench: Portable Tools for Performance Analysis. In USENIX annual technical conference (pp. 279-294).

\bibitem{dongarra2004}
Dongarra, J., Moore, S., Mucci, P., Seymour, K. and You, H., 2004. Accurate cache and TLB characterization using hardware counters. Computational Science-ICCS 2004, pp.432-439.

\bibitem{yotov2005}
Yotov, K., Pingali, K., \& Stodghill, P. (2005, September). X-ray: A tool for automatic measurement of hardware parameters. In Quantitative Evaluation of Systems, 2005. Second International Conference on the (pp. 168-177). IEEE.

\bibitem{yotov2005a}
Yotov, K., Pingali, K. and Stodghill, P., 2005. Automatic measurement of memory hierarchy parameters. ACM SIGMETRICS Performance Evaluation Review, 33(1), pp.181-192.

\bibitem{duchateau2008}
Duchateau, A.X., Sidelnik, A., Garzarán, M.J. and Padua, D., 2008, July. P-ray: A software suite for multi-core architecture characterization. In International Workshop on Languages and Compilers for Parallel Computing (pp. 187-201). Springer, Berlin, Heidelberg.

\bibitem{gonzalez2010}
González-Domínguez, J., Taboada, G.L., Fragüela, B.B., Martín, M.J. and Tourino, J., 2010, April. Servet: A benchmark suite for autotuning on multicore clusters. In Parallel \& Distributed Processing (IPDPS), 2010 IEEE International Symposium on (pp. 1-9). IEEE.

\bibitem{sandoval2011}
Sandoval, J.A., 2011. Foundations for Automatic, Adaptable Compilation (Doctoral dissertation, Rice University).


\bibitem{taylor2010}
Taylor, R. and Li, X.. A micro-benchmark suite for AMD GPUs. In Parallel Processing Workshops (ICPPW), 2010 39th International Conference on (387-396). IEEE.

\bibitem{sussman2011}
Sussman, A., Lo, N. and Anderson, T.. Automatic computer system characterization for a parallelizing compiler. In Cluster Computing (CLUSTER), 2011 IEEE International Conference on (pp. 216-224). IEEE.

\bibitem{abel2012measurement}
Abel, A., 2012. Measurement-based inference of the cache hierarchy (Doctoral dissertation, Master’s thesis, Saarland University, 2012.

 \bibitem{gonzalez2013s}
González-Domínguez, J., Martín, M.J., Taboada, G.L., Expósito, R.R. and Tourino, J., 2013. The Servet 3.0 benchmark suite: Characterization of network performance degradation. Computers \& Electrical Engineering, 39(8), pp.2483-2493.

\bibitem{casas2014active}
Casas, M. and Bronevetsky, G., 2014, May. Active measurement of memory resource consumption. In Parallel and Distributed Processing Symposium, 2014 IEEE 28th International (pp. 995-1004). IEEE.

\bibitem{casas2016evaluation}
Casas, M. and Bronevetsky, G., 2016. Evaluation of HPC applications’ memory resource consumption via active measurement. IEEE Transactions on Parallel and Distributed Systems, 27(9), pp.2560-2573.

\bibitem{moyer1991}
Moyer, S.A., 1991. Performance of the iPSC/860 node architecture. Institute for Parallel Computation, University of Virginia.

\bibitem{qasem2006}
Qasem, A. and Kennedy, K., 2006, June. Profitable loop fusion and tiling using model-driven empirical search. In Proceedings of the 20th annual international conference on Supercomputing (pp. 249-258). ACM.

\bibitem{luk2001}
Luk, C.K. and Mowry, T.C., 2001. Architectural and compiler support for effective instruction prefetching: a cooperative approach. ACM Transactions on Computer Systems, 19(1), pp.71-109.






\end{thebibliography}
\end{document}